\documentstyle[prl,aps,epsfig,multicol]{revtex}

\begin{document}
\title{Superconductivity in an organic insulator at very high magnetic fields.}
\author{L. Balicas$^1$, J. S. Brooks$^1$, K. Storr$^1$, S. Uji$^2$, M. Tokumoto$^3$, H. Tanaka$^4$, H. Kobayashi$^4$,
A. Kobayashi$^5$, V. Barzykin$^1$, and L. P. Gor'kov$^1$}
\address{$^1$National High Magnetic Field Laboratory, Florida State University, Tallahassee-FL 32306, USA.}
\address{$^2$National Research Institute for Metals, Tsukuba, Ibaraki 305-0003, Japan.}
\address{$^3$Electrotechnical Laboratory, Tsukuba, Ibaraki 305-8568, Japan.}
\address{$^4$Institute for Molecular Science, Okazaki, Aichi 444-8585, Japan}
\address{$^5$Research Centre for Spectrochemistry, Graduate School of Science, The University of Tokyo, Bunkyo-ku, Tokyo 113-0033, Japan}

\maketitle
\date{Received: \today }

\begin{abstract}
We investigate by electrical transport the field-induced superconducting state (FISC) in the organic conductor
$\lambda$-(BETS)$_2$FeCl$_4$. Below 4 K, antiferromagnetic-insulator, metallic, and eventually superconducting
(FISC) ground states are observed with increasing in-plane magnetic field. The FISC state survives between 18 and 41 T,
and can be interpreted in terms of the Jaccarino-Peter effect, where the external magnetic field {\em compensates}
the exchange field of aligned Fe$^{3+}$ ions. We further argue that the Fe$^{3+}$ moments are essential to stabilize the resulting
singlet, two-dimensional superconducting state
\end{abstract}


\begin{multicols}{2}
\narrowtext
Superconductivity is usually destroyed by diamagnetic currents induced in the presence of strong magnetic fields.
This effect has orbital character and prevails in most conventional ``s-wave" superconductors that involve singlet
state of the Cooper pairs. In addition, superconductivity can also be
suppressed by the Pauli pair breaking mechanism: here the external field destroys the spin-singlet state of the Cooper
pair, imposing the so-called Clogston-Chandrasekhar paramagnetic limit \cite{clogston,chandra}. Nevertheless, and despite these well
known physical limitations, S. Uji et al. \cite{uji} have recently reported the observation of a magnetic-field induced
superconducting phase (FISC) in the quasi-two-dimensional organic conductor $\lambda$-(BETS)$_2$FeCl$_4$ for fields
exceeding 18 tesla, applied parallel to the conducting layers. This is particularly remarkable since this compound,
at zero field, is an antiferromagnetic insulator (AI) below $T_p \cong 8.5 $K \cite{akoba}. The AI state is suppressed by
the application of magnetic fields above 10 tesla at low temperatures \cite{brossard}.

The present work was motivated by the apparent increase in the critical temperature of the FISC above 18 T
with increasing magnetic field (Ref. \cite{uji}). Here, for instance, in the case of spin-triplet
superconductivity, there would be in principle, no limit on the upper critical field.
The presence of Fe$^{3+}$ magnetic moments, which coexist with the FISC state, adds further appeal to the triplet
state model. To clarify the nature of the FISC, we have studied the $\lambda$-(BETS)$_2$FeCl$_4$ compound at
low temperatures in steady, tilted magnetic fields up to 42 tesla. Our main result is the observation of
reentrance towards the metallic state at a temperature-dependent critical field.
We obtain a temperature-magnetic field phase diagram for the FISC state,
which we interpret in terms of the Jacarino-Peter (JP) field compensation effect \cite{jaccarino}. This implies that the Cooper
pairs condense into a spin-singlet state. We argue further that the Fe$^{3+}$ magnetic state is indeed necessary to
stabilize the singlet superconducting state by suppression of diamagnetic currents in the associated in-plane high
magnetic fields.

$\lambda$-(BETS)$_2$FeCl$_4$ (where BETS stands for Bis(ethylenedithio)tetraselenafulvalene) crystallizes in
a triclinic unit cell. The BETS planar molecules are stacked along the crystallographic a-axis, and constitute
conducting planes parallel to the a-c plane. These conducting layers alternate along the b-axis with layers
containing linear chains of FeCl$_{4}^{-}$ magnetic anions, hence the b-axis is the least conducting direction.
Spin interactions between localized Fe$^{3+}$ 3d electrons and $\pi$ conducting electrons are expected due to
the short interatomic distance between the BETS molecules and the FeCl$_4$ anions.

Single crystals of $\lambda$-(BETS)$_2$FeCl$_4$ were obtained by electro-crystallization \cite{hkoba}.
Annealed (low strain) gold wires ($\phi$ = 12.5 $\mu$m) were attached with graphite paint in a four-terminal
arrangement along the c-axis. An ac current (10 to 100 $\mu$A) was used, and the voltage was measured by a
conventional lock-in amplifier technique. Samples were mounted in a rotating sample holder in a $^3$He refrigerator.
The measurements were carried out in the Hybrid magnet at the DC Field Facility of the National High Magnetic Field
Laboratory.

Our magnetic field dependent resistance of a $\lambda$-(BETS)$_2$FeCl$_4$ single crystal is shown in Fig. 1(a) for
different temperatures. Here the magnetic field $B$ is applied along the in-plane c-axis.
The main characteristic of the data is that between 18 and 41 tesla, the resistance of the material drops
with decreasing temperature, reaching zero within experimental uncertainties below 2 K in a field range centered
near 33 T. It is important to mention that in this part of the ($B-T$) phase diagram, the material behaves as a
good metal. We find Shubnikov-de Haas oscillations (of order 700 T with effective mass $m^{\star} \sim 4m_{0}$) for
the magnetic field perpendicular to the conducting planes \cite{ujibalicas}, which for an isotropic model would give a
Fermi energy $\varepsilon_{F} \sim$ 200 K. The normal state resistivity is of order of 10$^{-4}$ $\Omega$.cm in
the metallic state near 15 T. We estimate $k_{F} \ell \sim 20$ (where $k_{F}$ is the Fermi wave vector and $\ell$ is the mean free path)
and thus, despite the low scale of $\varepsilon_{F}$, the standard metallic conditions are fulfilled.

In the FISC state at higher fields, the resistivity drops typically by 2 to 4 orders of magnitude,
putting it at or below the conductivity of copper, and beyond our ability to measure by standard ac lock-in methods.
From the isothermal field scans we can extract the temperature dependence of the resistance at fixed values of the
field, see Fig. 1(b). For fields between 18 and 37 T, the resistance shows a phase transition from the metallic phase
(above 4.2 K) to the full superconducting state. Above a certain threshold field $B_{th}=$18 tesla, the onset of this transition
{\em increases} with magnetic field, reaching a maximum $T_{c} \cong 4$ K at $B^{\star}$ = 33 tesla. Above $B^{\star}$
the onset {\em decreases} in temperature with increasing field, and above 41 T the FISC is suppressed rapidly. We note that
the experimental resistance does not fall immediately to zero below $T_{c}$. We expect, given the very small,
delicate nature of the samples, that the presence of strain, sample quality, or sample mis-alignment may cause
some inhomogeneity (percolation) in the superconducting fraction at the onset of the FISC.

We next discuss a central question concerning the interpretation of the FISC state as ``truly" superconducting,
beyond the observation of zero resistance within experimental uncertainties. The Meissner effect - the standard
test for the onset of superconductivity - where magnetic flux is excluded when a sample enters the
superconducting state, may become a non-trivial experiment in the present case. This is due to the fact that
the magnetic flux may be trapped between two-dimensional superconducting layers. 
However, as torque magnetization measurements show \cite{uji}, there is a
{\em bulk} phase transition at $B_{th}(T)$ from the metallic to the FISC state.

The present work provides two additional, independent pieces of evidence that the FISC is superconducting.
The first is that the state is re-entrant to a metallic state above 41 T what excludes triplet pairing. 
This observation also rules out field-induced low-resistance models. When magnetic scattering, or some other form of higher 
resistance state is removed by magnetic fields, restoration of disorder-related, inelastic processes at higher fields is 
very unlikely. The second observation is the effect of the transverse magnetic field (i.e. field perpendicular to the layers) 
on the FISC state. This behavior is illustrated in Fig. 2(a), where we show results from a systematic variation of the magnetic 
field away from the in-plane orientation at the lowest temperature of our investigation. The essential detail here is that the zero
resistance state begins to vanish for a {\em transverse field} $B_{c \bot}$ greater than 3.5 T. This observation is elucidated in
Fig. 2(b) by plotting the resistance for a {\em constant in-plane field} $B_{c \|} = B sin(\theta)$ of about 33 T
(i.e. $B_{c \|}  = B^{\star}$ ) vs. the transverse field $B_{c \bot} = B cos(\theta)$. Hence the FISC state is
removed when orbital components appear. The most striking observation is that the critical field $B_{c \bot}$ for
the FISC state is essentially identical to that of the non-magnetic, isostructural material $\lambda$-(BETS)$_2$GaCl$_4$
\cite{tanatar}, and by comparison, this suggests that the FISC is also a singlet superconducting state.

We next consider how the FISC state is stabilized. While the two anions (Fe$^{3+}$ and Ga$^{3+}$) have different ground
states at $T = 0$ in the low field range of the ($B-T$) phase diagram, alloying by Ga and external pressure
restores superconductivity in the Fe-based material \cite{sato}. We expect that the superconducting states, in both cases,
are close in energy. We therefore argue that the in-plane physics is similar for both materials, and the differences
in the phase diagrams arise from correspondingly small energies related to, for instance, the inter-layer coupling.
Our model is as follows. In-plane fields orient the $S = 5/2$ spin of the Fe$^{3+}$ ions, and we assume that this
decouples the BETS conducting layers. The problem then becomes two-dimensional (2D), with no diamagnetic currents
flowing between layers in the presence of a purely in-plane magnetic field (we will return to this point later).
For the 2D geometry, the in-plane field can destroy singlet superconductivity by means of paramagnetic effects only,
i.e., breaking the Cooper pairs \cite{clogston}. The $S = 5/2$ Fe$^{3+}$ magnetic moments, oriented by magnetic field, exert the
exchange field $J \langle S \rangle$ on the spins $s$ of the conduction electrons via the exchange interaction,
$J \vec{s} \cdot \vec{S}$. Thus, the effective field $H_{eff}$ acting on the electron spin is:
\begin{equation}
I(B) = \mu _{B} H_{eff} = \mu _{B} B + J \langle S \rangle
\end{equation}
Eq. (1) is at the heart of the Jaccarino-Peter effect \cite{jaccarino}: for $J<0$ the two contributions will compensate each other
to restore superconductivity. For our experiments with $B \cong 20$ - 40 tesla and $T < 5$ K, the iron moments are saturated:
$\langle S \rangle = 5/2$. Following Refs. \cite{clogston,jaccarino}, the restoration of superconductivity (at $T = 0$) will occur in the field
interval $|I(B)| < 0.755 \Delta(0)$, where $\Delta(0)$ is the superconducting gap at $T = 0$ (in the absence of the field).
The situation, however, is more complicated by the fact that the phase transitions separating normal and
superconducting states in the $(T, I(B))$ - plane, may be either 1$^{st}$ or 2$^{nd}$ order \cite{lev}.
The phase diagram has recently been revised by two of us in \cite{victor} (see also \cite{balents}); we plot in Fig. 3 the theoretical
results, together with the experimental data in the ($T, B$)-plane. Experiments suggest $I(B^{\star}) = 0$ at $B^{\star} = 33$ tesla,
$T_{c} \cong 4.2$ K, and these parameters were used in the theoretical plots in Fig. 3.  In Fig. 3 the solid line is
the 2$^{nd}$ order transition line. At lower temperatures, transitions from the normal to the superconducting states begin with
a 2$^{nd}$ order transition into the inhomogeneous LOFF-state (shaded area) \cite{loff}, rapidly followed by a 1$^{st}$ order
transition (dashed curve) into the homogeneous state. The branching points are at $T_{t} \cong 2.3$ K. The
range of existence of the LOFF state is rather narrow \cite{victor}: we estimate the width (at $T = 0$) as only $\Delta B \cong
0.72$ T. (In addition, this state is very sensitive to defects). We see from Fig. 3 that the theoretical model
discussed above reproduces the main features of the FISC phase diagram very well.

Let us now return to the question regarding the nature of the new FISC state in $\lambda$-(BETS)$_2$FeCl$_4$ and its
relation to the ``conventional" SC state in $\lambda$-(BETS)$_2$GaCl$_4$. Here we argue that remarkably, the Fe$^{3+}$
magnetic ions may be essential to stabilize the two-dimensionality of the  FISC state produced by the JP effect.
Both compounds have similar, anisotropic, layered structure. Nevertheless, there is an inter-plane electronic
coupling in the $\lambda$-(BETS)$_2$GaCl$_4$, as is evidenced in the finite upper critical field \cite{tanatar} for the
in-plane direction, $H_{C2}(0)= 12-15$ tesla, which is well below the field where the FISC is stable. To restore
superconductivity in $\lambda$-(BETS)$_2$FeCl$_4$ within the field range of the JP compensation effect, one
needs a mechanism to fully eliminate the diamagnetic interlayer currents in this compound. We suggest that the
coupling between 2D BETS layers comes about through the bridging of MCl$_4$–tetrahedra, so that in the second order
effective tunneling matrix element, the (MCl$_4$)$^0$-state shows up as the intermediate state in simple perturbation theory.
While the p-shell is empty for Ga$^{3+}$, all the ``d-levels" are occupied in Fe$^{3+}$, according to Hund's rule.
Placing an additional electron at the Fe$^{3+}$ site (with spin antiparallel to $\vec{S}$) is energetically unfavorable,
i.e. it costs the Hund coupling energy. Furthermore, when the $S = 5/2$ spins of Fe$^{3+}$ are aligned by the field,
an electron with parallel spin has no accessible states on the d-shell, and the energy cost inside the FeCl$_4$-complex
is expected to increase even further. Therefore, tunneling across magnetically oriented Fe$^{3+}$-tetrahedra becomes a spin selective
process, consequently, transport of an s-wave Cooper pair between adjacent planes is excluded. 
Clearly, in $\lambda$-(BETS)$_2$GaCl$_4$ there is no spin selective process, and hence a correspondingly smaller upper 
critical field is observed \cite{tanatar}.

Finally, we consider why $T_{c}^{\star} \cong 4.2$ K in $\lambda$-(BETS)$_2$FeCl$_4$ is less than
$T_{c}^{\star} \cong 5.5$ K for $\lambda$-(BETS)$_2$GaCl$_4$. Our guess is that both superconducting ground states
(due to the two-dimensionality) are basically the same at $T = 0$, (i.e. $\Delta(0)_{Fe} \cong \Delta(0)_{Ga}= 1.76 T_{c}^{Ga}$).
$T_{c}^{\star} = T_{c}^{Fe}$ seems to be smaller, since at higher temperatures, thermal activation of the
Fe$^{3+}$ spins via the exchange interaction $J$ provides a mechanism for pair breaking scattering.

We conclude by noting that $\lambda$-(BETS)$_2$FeCl$_4$ (along with the non-magnetic analog $\lambda$-(BETS)$_2$GaCl$_4$) has
provided a rich new area for the study of low dimensional superconductivity and magnetism, where the two mechanisms
compete on a very low energy scale. In this paper we have provided a simple theoretical picture that describes
the broader features of the newly discovered high field induced superconducting state. We show experimentally that 
in-plane diamagnetic currents ($B_{c \bot}$) can destroy the FISC state. We have argued that magnetic ions are actually
essential to suppress the coupling between planes in the presence of in-plane magnetic fields. 
However, there are many unusual features in this system that will require a significant, further level of both 
experimental and theoretical work. In particular, the mechanism by which the magnetic field penetrates in the bulk
of our 2D superconducting samples remains unclear and may produce hysteretic behavior.

We would like to thank the Hybrid Magnet Group at the NHMFL for their
invaluable assistance during these measurements. One of us (JSB) acknowledges support
from NSF-DMR-99-71474 for this work. The NHMFL is supported by a cooperative agreement between the
State of Florida and the National Science Foundation through NSF-DMR-0084173.

\begin{figure}[h]
\begin{center}
\epsfig{file=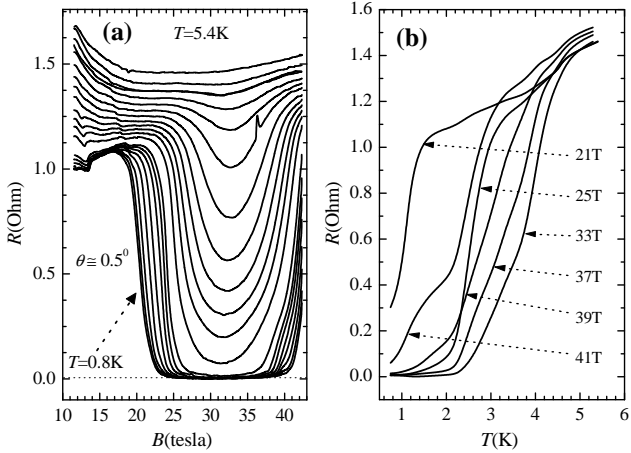, width=8.6cm}
\caption{(a) Resistance $R$ as a function of magnetic field $B$, applied along the 
in-plane c-axis ($\pm 0.3$ degrees) of a $\lambda$-(BETS)$_2$FeCl$_4$ single 
crystal for temperature intervals of approximately 0.25 K, between 5.4 and 0.8 K.
The superconducting state develops progressively with decreasing temperature, but 
is suppressed for fields sufficiently away
from (above or below) 33 tesla.  (We note that since the Hybrid magnet is composed 
of a superconducting outsert coil in combination
with a Bitter type resistive insert coil, the field generated by the outsert is
kept constant at approximately 11.5 tesla,
while the field of the insert coil was ramped between 0 and and 31.5 tesla). (b)
 Resistance as a function of temperature
$T$ for several values of $B$ obtained from the field scans shown in (a). The FISC 
transition has a maximum transition
temperature $T_{c} \simeq 4.2$ K near 33 tesla.}
\end{center}
\end{figure}
\begin{figure}[htpb]
\begin{center}
\epsfig{file=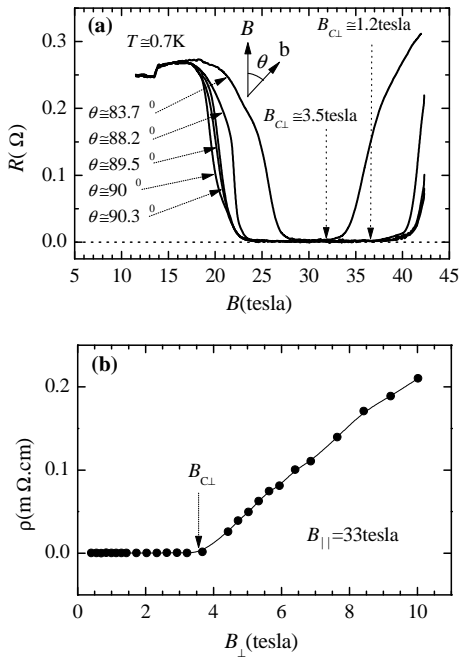, width=7cm}
\caption{(a) Resistance as a function of magnetic field at $T = 0.7$ K and for five
different angles $\theta$ (indicated in the figure) between $B$
and the inter-plane b-axis. Notice that the inter-plane critical
field $B_{c \bot}$, defining the orbital effect, decreases as
$\theta$ approaches 90$^ \circ$. (b) Resistance for constant
in-plane field $B_{c \|}$ vs transverse magnetic field $B_{c \bot}$
at $T = 0.7 $ K. }
\end{center}
\end{figure}
\begin{figure}
\begin{center}
\epsfig{file=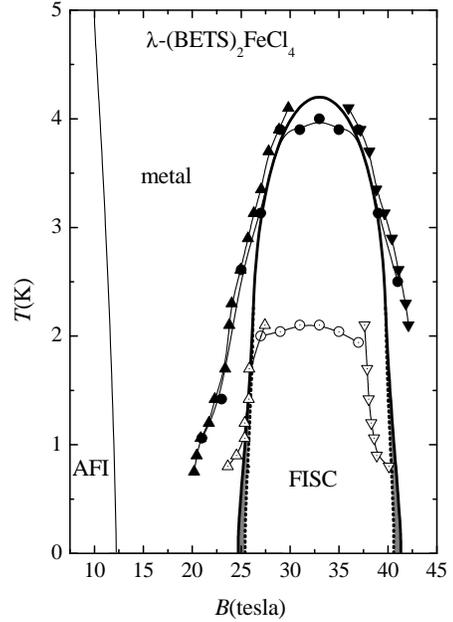, width=7cm}
\caption{Temperature-magnetic field phase diagram showing the AFI, metallic, and
 FISC states for a $\lambda$-(BETS)$_2$FeCl$_4$ single crystal vs in-plane 
magnetic field. Solid triangles indicate the middle point of the resistive transition as a
function of $B$ (from Fig. 1A), while solid circles indicate the middle point of
the resistive transition as a function of $T$ (from Fig. 1B). Open triangles and circles indicate the region
where the resistance vanishes at the level of sensitivity of our instrumentation.
The solid line is a theoretical fit (see text) to a second order phase transition 
towards the FISC while the dashed line indicates a first order transition from the 
inhomogeneous LOFF state (shaded area) into the bulk SC state.}
\end{center}
\end{figure}

\end{multicols}

\begin{references}

\bibitem{clogston}  A. M. Clogston, Phys. Rev. Lett. {\bf 9}, 266 (1962).
\bibitem{chandra} B. S. Chandrasekhar, Appl. Phys. Lett. {\bf 1}, 7 (1962).
\bibitem{uji} S. Uji, H. Shinagawa, C. Terakura, T. Terashima, T. Yakabe {\em et al.}, Nature (in press).
\bibitem{akoba} A. Kobayashi, T. Udagawa, H. Tomita, T. Naito, H. Kobayashi, Chem. Lett., 2179 (1993).
\bibitem{brossard} L. Brossard {\em et al.}, Eur. Phys. J. B {\bf 1}, 439 (1998).
\bibitem{jaccarino} V. Jaccarino and M. Peter, Phys. Rev. Lett. {\bf 9}, 290 (1962).
\bibitem{hkoba} H. Kobayashi, H. Tomita, T. Naito, A. Kobayashi, F. Sakai, T. Watanabe, and P. Cassoux,
J. Am Chem. Soc. {\bf 118}, 368 (1996).
\bibitem{ujibalicas} S. Uji {\em et al.}, to be published;  L. Balicas {\em et al.}, to be published.
\bibitem{tanatar} M. A. Tanatar, T. Ishiguro, H. Tanaka, A. Kobayashi and H. Kobayashi, J. of Supercond. {\bf 12}, 511 (1999).
\bibitem{sato} A. Sato, E.  Ojima, H. Akutsu, Y. Nakazawa, H. Kobayashi, H. Tanaka, A. Kobayashi, and P. Cassoux, Phys. Rev. B
{\bf 61}, 111 (2000).
\bibitem{lev} L. P. Gor'kov and A. I. Rusinov, Sov. Phys. JETP {\bf 19}, 922 (1964).
\bibitem{victor} V. Barzykin and L. P. Gor'kov, Phys. Rev. Lett. {\bf 89}, 2207 (2000).
\bibitem{balents}  L. Balents and C. M. Varma, Phys. Rev. Lett. {\bf 89}, 1264 (2000).
\bibitem{loff}  A. I. Larkin and Yu. N. Ovchinnikov, Sov. Phys. JETP {\bf 20}, 762 (1962);
P. Fulde and R. A. Ferrell, Phys. Rev. {\bf 135A}, 550 (1964).
\end{references}
\end{document}